\documentclass[aps,prl,reprint,superscriptaddress]{revtex4-2}
\usepackage{graphicx}
\usepackage{bm,amssymb}
\usepackage{amsmath}
\usepackage{graphics,subcaption}

\usepackage{pdfpages} 
\usepackage{pgffor} 

\makeatletter
\AtBeginDocument{\let\LS@rot\@undefined}
\makeatother

\def\supplementfilename{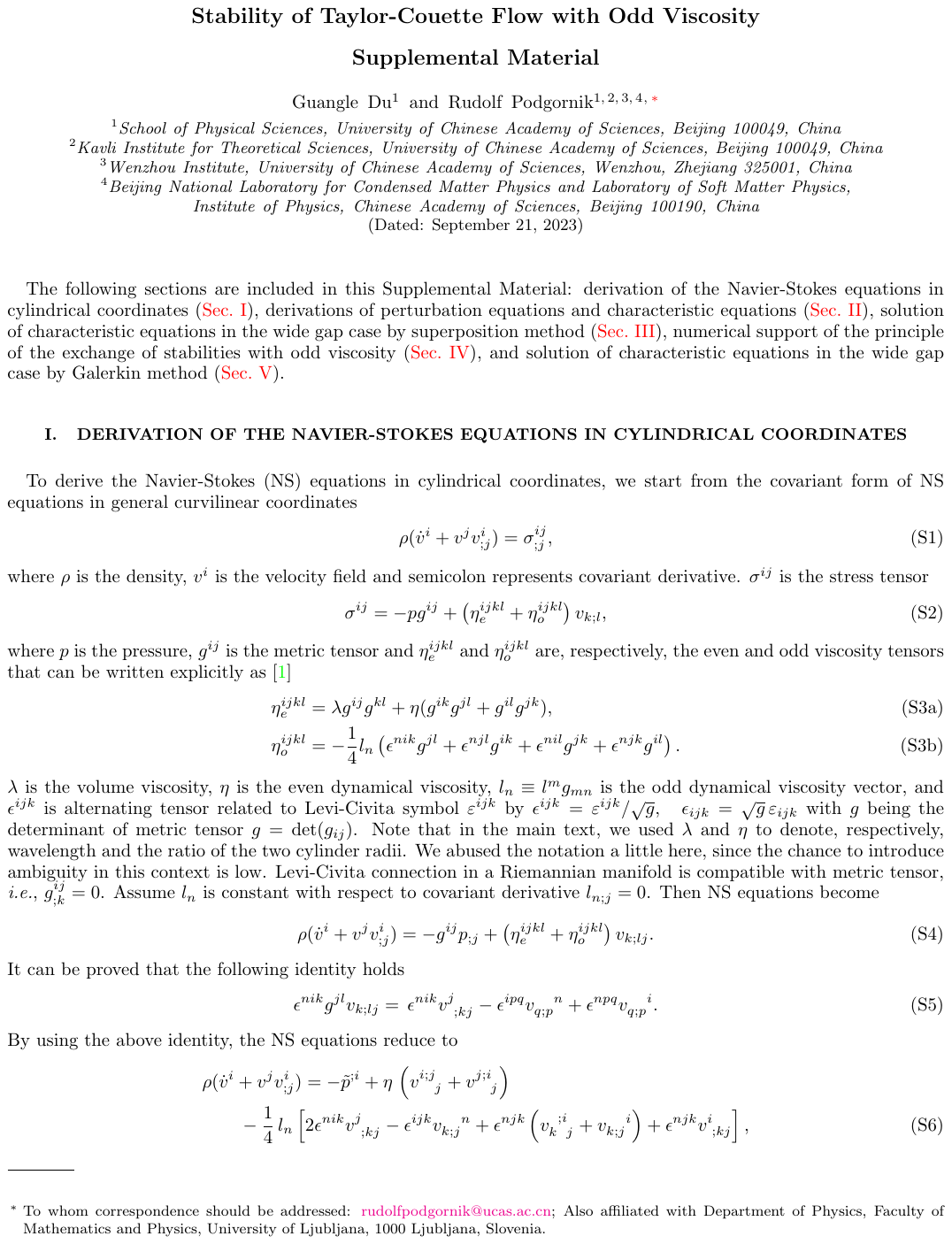}

\pdfximage{\supplementfilename}
\def\numbersupplementpages{\the\pdflastximagepages}

\newif\ifarXiv
\arXivtrue

\begin{document}

\title{Stability of Taylor-Couette Flow with Odd Viscosity}
\date{\today}
\author{Guangle Du}
\affiliation{School of Physical Sciences, University of Chinese Academy of Sciences, Beijing 100049, China}

\author{Rudolf Podgornik}
\email[To whom correspondence should be addressed: ]{rudolfpodgornik@ucas.ac.cn}
\altaffiliation[Also affiliated with ]{Department of Physics, Faculty of Mathematics and Physics, University of Ljubljana, 1000 Ljubljana, Slovenia.}
\affiliation{School of Physical Sciences, University of Chinese Academy of Sciences, Beijing 100049, China}
\affiliation{Kavli Institute for Theoretical Sciences, University of Chinese Academy of Sciences, Beijing 100049, China}
\affiliation{Wenzhou Institute, University of Chinese Academy of Sciences, Wenzhou, Zhejiang 325001, China}
\affiliation{Beijing National Laboratory for Condensed Matter Physics and Laboratory of Soft Matter Physics, Institute of Physics, Chinese Academy of Sciences, Beijing 100190, China}

\begin{abstract}
Odd viscosity can emerge in 3D hydrodynamics when the time reversal symmetry is broken and anisotropy is introduced.
Its ramifications on the stability of the prototypical Taylor-Couette flow in curved geometries have remained unexplored. Here, we investigate the effects of odd viscosity on the stability of Taylor-Couette flow under axisymmetric perturbations both analytically and numerically, deriving analytically the critical Taylor number for different odd viscosities in the narrow gap case as well as fully numerically implementing the wide gap case. We find that the odd viscosity modifies the vortex pattern by creating secondary vortices, and exerts an intriguing ``lever'' effect in the stability diagram, completely suppressing the instability of Taylor-Couette flow under axisymmetric perturbations when the odd viscosity is large, irrespective of its sign. Our findings highlight the role of odd viscosity for the rich flow patterns of the Taylor-Couette geometry and provide guidance for viscometer experiments when odd viscosity is present.
\end{abstract}
\maketitle
\paragraph{Introduction.}
The viscosity tensor, connecting viscous stress with deformation rate in Newtonian hydrodynamics with microscopic time reversal symmetry, is subject to the Onsager reciprocal relations~\cite{onsagerReciprocalRelationsIrreversible1931a}, being a symmetric tensor with respect to the major indices~\cite{landauFluidMechanics1987,guyonPhysicalHydrodynamics2015}.
An anti-symmetric and consequently non-dissipative part of the viscosity tensor, referred to as the odd (or Hall/gyro) viscosity~\cite{avronOddViscosity1998, fruchartOddViscosityOdd2023},
can emerge when Onsager reciprocal relations are violated due to the breaking of the time reversal symmetry either spontaneously as in, \emph{e.g.},
superfluids~\cite{vollhardtSuperfluidPhasesHelium1990,hoyosEffectiveTheoryChiral2014} and superconductors~\cite{shitadeBulkAngularMomentum2014,roseHallViscosityConductivity2020}, or explicitly by external magnetic fields or global/local rotations in systems such as quantum Hall liquids~\cite{offertalerViscoelasticResponseQuantum2019,offertalerViscoelasticResponseQuantum2019},
polyatomic gases~\cite{knaapHeatConductivityViscosity1967,korvingInfluenceMagneticField1967}, magnetized plasma~\cite{thompsonDynamicsHighTemperature1961,robertsMagnetohydrodynamicEquationsFinite1962a},
rotated gases~\cite{hooymanCoefficientsViscosityFluid1954,nakagawaKineticTheoryGases1956} and chiral active fluids~\cite{banerjeeOddViscosityChiral2017,hanFluctuatingHydrodynamicsChiral2021}.
The form of odd viscosity is not universal but is in addition constrained by spatial dimension and symmetry of the fluid.
While in the 2D case odd viscosity is compatible with isotropy and its effects have been studied extensively~\cite{avronOddViscosity1998,banerjeeOddViscosityChiral2017,ganeshanOddViscosityTwodimensional2017,souslovTopologicalWavesFluids2019,louOddViscosityinducedHalllike2022,fruchartOddViscosityOdd2023}, in the more realistic 3D case odd viscosity can only survive under lower symmetries than isotropy, \emph{e.g.}, cylindrical symmetry~\cite{markovichOddViscosityActive2021,khainStokesFlowsThreedimensional2022}.

Of the various consequences imparted by the odd viscosity in 3D such as anisotropic bulk shear waves~\cite{markovichOddViscosityActive2021}, generation of transverse flow past spheres or bubbles~\cite{khainStokesFlowsThreedimensional2022}, Taylor column and inertial waves~\cite{kirkinisTaylorColumnsInertiallike2023,kirkinisInertiallikeWavesRigidlyrotating2023}, \emph{etc}.,
one particular ramification demanding in-depth investigations is its effects on hydrodynamic instabilities, which determine the conditions of whether flows with odd viscosity can actually survive in the real world. Instabilities, including Rayleigh–Taylor and Kelvin–Helmholtz, have been investigated in the context of magnetized plasma, where it was found that the odd viscosity can either have a stabilizing or  a destabilizing effect, depending on its sign~\cite{nayyarEffectGyroviscosityRayleighTaylor1970,wolffOnsetDevelopmentKelvinHelmholtz1980,rudenPolarityDependentEffect2004}.
In the context of liquid films in plane geometries, the presence of odd viscosity tends to suppress, {\sl e.g.},  thermocapillary, Saffman–Taylor and Faraday instabilities~\cite{kirkinisOddviscosityinducedStabilizationViscous2019,baoOddviscosityinducedInstabilityFalling2021,chuEffectOddViscosity2022,reynoldsHeleShawFlowParity2022,samantaRoleOddViscosity2022,chuElectrostaticallyInducedFaraday2023}.
In this context it therefore comes as a surprise that the effect of odd viscosity on the prototypical instability in curved geometries, the {\sl Taylor-Couette instability}, remains virtually unexplored.

The Taylor-Couette instability occurs in the flows between concentric rotating cylinders at a critical adverse distribution of angular velocity~\cite{taylorStabilityViscousLiquid1923}, and is paradigmatic for the studies of hydrodynamic stabilities and transitions in curved geometries~\cite{Di-Prima-1985,charruHydrodynamicInstabilities2011,fardinHydrogenAtomFluid2014}.
The simple geometry can give rise to a rich series of distinct flows with increasing complexity that finally lead to fully developed  turbulence~\cite{andereckFlowRegimesCircular1986}.
In addition, the instability of centrifugal origin is generic and closely related to other instabilities of the flows deflected by curved boundaries such as Dean and G\"{o}rtler vortices~\cite{charruHydrodynamicInstabilities2011,guyonPhysicalHydrodynamics2015}.
Moreover, experimentally the Taylor-Couette setting has been widely used in the context of viscometry ever since the first viscometer was build by Mallock~\cite{mallockDeterminationViscosityWater1889} and Couette~\cite{couetteNouvelAppareilPour1888}.

In this Letter, we investigate comprehensively how the presence of odd viscosity changes the Taylor-Couette instability under axisymmetric perturbations in both the narrow as well as the wide gap cases. By both analytical and numerical calculations we found that odd viscosity can qualitatively change the nature and the stability of the flow, \emph{i.e.}, the instability under axisymmetric perturbations can be completely suppressed by large odd viscosity, irrespective of its sign. Our findings not only illuminate the effects of odd viscosity on the stability of Taylor-Couette flow, but also offer clues as to its more complex patterns and provide guidance for experiments investigating the odd viscous flows in curved geometries.

\paragraph{Setups.}

\begin{figure}[ht]
  \centering
  \includegraphics[width=0.9\linewidth]{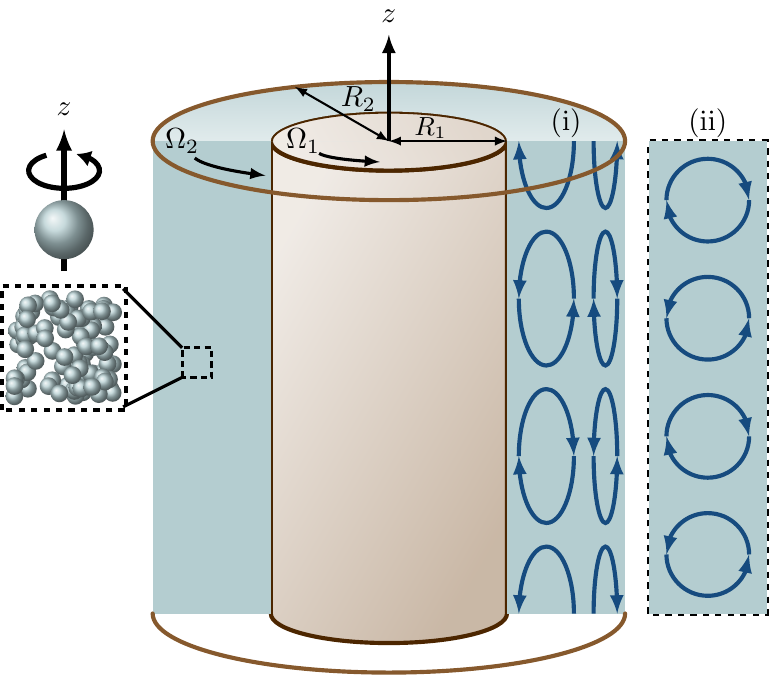}
  \caption{\label{fig:schematic}Schematics of the Taylor-Couette flow with odd viscosity, resulting in deformed Taylor vortices and secondary vortices near the outer surface (i), in contrast to the ordinary flow \emph{without} odd viscosity, leading to approximately square vortices with a single vortex along radial direction (ii).}
\end{figure}

As shown in Fig.~\ref{fig:schematic}, we consider the Taylor-Couette flow confined in two infinite concentric rotating cylinders with radii \(R_1\) and \(R_2\) and angular velocities \(\Omega_1\) and \(\Omega_2\), respectively. The gap between the two cylinders is denoted by \(d = R_2 - R_1\). The \(z\)-direction is fixed to be along that of \(\Omega_1\). The spinning motion of fluid molecules breaks the time-reversal symmetry and hence induces an odd viscosity.
A study starting from microscopic theory shows that the odd viscosity in this case is proportional to the density of molecular angular velocity~\cite{markovichOddViscosityActive2021}.
For simplicity, we assume that only the \(z\)-component of the odd viscosity vector is nonvanishing and denote it as \(\nu_o\). We adopt the Boussinesq approximation where the density \(\rho\), even and odd kinematic viscosities \(\nu,\, \nu_o\) are approximately constant.
Denote the velocity field in cylindrical coordinates as \(\mathbf{u} = (u_r,\,u_{\theta},\,u_z)\).
The Navier-Stokes (NS) equations for the Taylor-Couette flow can then be explicitly written as~\cite{markovichOddViscosityActive2021,suppl}
\begin{subequations}
\begin{align}
&D_tu_r\! -\! \frac{u_{\theta}^2}{r} = \nu\! \left(\! \Delta^{*} u_r - \frac{2 \partial_{\theta} u_{\theta}}{r^2} \!\right) + \frac{\nu_o}{2} \partial_z \omega_r- \frac{\partial_r \tilde{p}}{\rho}, \\
&\left(\! D_t \!+\! \frac{u_r}{r}\!\right)\! u_{\theta} =  \nu \! \left(\! \Delta^{*} u_{\theta} + \frac{2\partial_{\theta} u_r }{r^2}\!\right)\!  + \! \frac{\nu_o}{2} \partial_z \omega_\theta\!-\! \frac{\partial_{\theta} \tilde{p}}{\rho \, r}, \\
& D_t u_z = \nu \Delta u_z +\frac{\nu_o}{2} \partial_z \omega_z - \frac{\partial_z \tilde{p}}{\rho},
\end{align}
\end{subequations}
where \(\omega_r = \left( \partial_{\theta} u_z/r - \partial_z u_{\theta} \right)/2\),
\(\omega_{\theta} = \left( \partial_z u_r - \partial_r u_z \right)/2\) and
\(\omega_z = \left( \partial_r u_{\theta} + u_{\theta}/r - \partial_{\theta} u_r/r \right)/2\) are, respectively, the components of the vorticity vector and \(\tilde{p} = p + \nu_o\rho\,\omega_z\) is an effective pressure. Additionally we introduced \(D_{t} = \partial_t + u_r \partial_r + u_{\theta}\partial_\theta /r + u_z \partial_z\), \(\Delta = \partial^2_r + \partial_r/r +  \partial^2_{\theta}/r^2+ \partial^2_z\) and \(\Delta^{*} = \Delta - 1/r^{2}\).
We assume the flow is incompressible and the equation of continuity becomes \(\partial_r u_r + u_r/r + \partial_{\theta} u_{\theta}/r + \partial_z u_z = 0\).
Below we will investigate how the presence of the odd viscosity \(\nu_o\) changes the stability of the Taylor-Couette flow by performing a linear stability analysis and obtaining the stability diagrams for both narrow and wide gap cases. We find that depending on the sign of odd viscosity, the instability can be either promoted or suppressed. What is remarkable is that when the absolute value of odd viscosity is large enough, the instability can {\em completely disappear}.

\paragraph{Stationary solution and perturbation equations.}
The NS equations with the odd viscosity allow for the stationary solution of the form
\(u_r = u_z = 0\), \(u_{\theta} = V(r) = r \Omega(r)\), and \(\tilde{p} = \tilde{P}(r)\).
The azimuthal velocity \(V\) can be obtained by solving
\(\frac{\mathrm{d}^2 V}{\mathrm{d} r^2} + \frac{1}{r} \frac{\mathrm{d} V}{\mathrm{d}r} - \frac{V}{r^2} = 0\)
along with the no-slip boundary conditions on the surfaces of the cylinders, namely,~\cite{chandrasekhar1961,drazinHydrodynamicStability2004}
\begin{align}
\label{eq:stationary}
V(r) = A r + \frac{B}{r}, \,
A = \Omega_1 \frac{\mu - \eta^2 }{1-\eta^2}, \, B = \Omega_1 R_1^2 \frac{1-\mu}{1-\eta^2},
\end{align}
where \(\eta=R_1/R_2\) and \(\mu=\Omega_2/\Omega_1\).

From the case without the odd viscosity, we know that it is the axisymmetric perturbations that dominate at the initial stage of the instability~\cite{chandrasekhar1961,drazinHydrodynamicStability2004}.
Hence we consider axisymmetric perturbations characterized by
\(\mathbf{u} = (u'_r,\, V + u'_{\theta},\, u'_z)\) and \(\tilde{p} = \tilde{P} + \tilde{p}'\).
We then perform the standard normal mode analysis
\((u'_r,\, u'_{\theta},\, u'_z) = (u(r),\,v(r),w(r)) \mathrm{e}^{s t +\mathrm{i} kz}\) and \(\tilde{p}' = q(r)\, \mathrm{e}^{s t +\mathrm{i} kz}\).
The marginal state of the instability can either be stationary \(s = 0\) or oscillating with \(s\) being purely imaginary. The principle of the exchange of stabilities stipulating the existence of a stationary marginal state in the case without odd viscosity was verified experimentally~\cite{taylorStabilityViscousLiquid1923,chandrasekhar1961}.
We now assume that this principle still holds in the presence of the odd viscosity, and will use numerical calculations to support it later~\cite{suppl}.

Next, define the Taylor number and the reduced odd viscosity as~\cite{alternativeDefTaylorNumber}
\begin{align}
T \equiv - \frac{4A\Omega_1 d^4}{\nu^2},\quad \bar{\nu}_o \equiv \frac{\nu_o}{8\Omega_1 d^2}.
\end{align}
By using \(\zeta = (r - R_1)/d\) and \(a = kd\) and substituting \(u \to 2\Omega_1 d^2 a^2\, u/\nu\), we have the
characteristic equations
\begin{subequations}
\begin{align}
\label{eq:wide-char1}
&\left( DD_{*} - a^2\right)^2 u = \left[ g - \bar{\nu}_o \left( DD_{*} - a^2 \right) \right]v, \\
\label{eq:wide-char2}
& \left( DD_{*} - a^2\right) v = - T a^2 \left[ 1 - \frac{\bar{\nu}_o}{f}  \left( DD_{*} - a^2 \right) \right] u,
\end{align}
\end{subequations}
where we have denoted
\(D = \frac{\mathrm{d} }{\mathrm{d}\zeta }\), \(D_{*} = \frac{\mathrm{d} }{\mathrm{d}\zeta } + 1/[\zeta + \eta/(1-\eta)]\),
\(f(\eta,\mu) = (\mu - \eta^2 )/(1- \eta^2)\) and
\(g(\eta,\mu,\zeta) = \frac{\mu - \eta^2}{ 1 -\eta^2} + \frac{1-\mu}{1-\eta^2} \left( 1 + \frac{1-\eta}{\eta} \zeta \right)^{-2}\).
The no-slip boundary conditions stipulate that $u = v = Du = 0$ at $\zeta=0$ and $\zeta=1$.

\begin{figure}[t]
  \centering
  \includegraphics[width=0.99\linewidth]{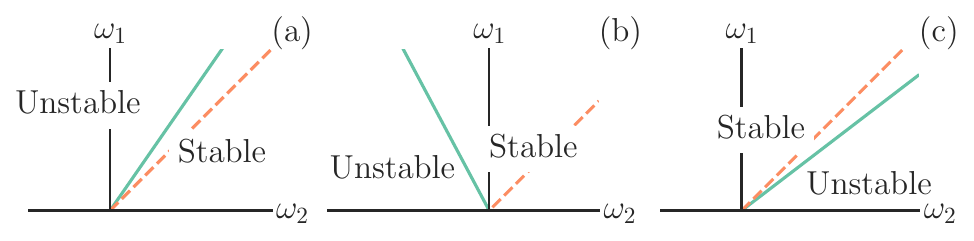}
  \caption{\label{fig:phase-diagram-inviscid}Stability diagrams in the $(\omega_2,\, \omega_1)$ plane in the inviscid case. $\omega_{i}= \Omega_i R_i^2/\nu,\, i=1,2$. The solid lines separating the stable and unstable regions represent the critical conditions for different odd viscosities, while the dashed lines represent the Rayleigh criterion $\omega_1=\omega_2$ without odd viscosity. (a) $0 < \bar{\nu}_o < \bar{\nu}_{\text{oc}}$; (b) $\bar{\nu}_{o} > \bar{\nu}_{\text{oc}}$; (c) $\bar{\nu}_{o} < 0$.}
\end{figure}

\paragraph{Inviscid case.}
We first consider the simple inviscid case with \(\nu\to 0\).
The reduced NS equations along with the boundary conditions \(u_r=0\) at \(r=R_1\) and \(r=R_2\) allow a stationary solution of the form
\(u_r = u_z = 0\), \(u_{\theta} = V(r) = r \Omega(r)\) and \(\tilde{p} = \tilde{P}(r)\), where
\(V(r)\) can be an arbitrary function of \(r\).
Here we focus on \(V(r)\) of the form in Eq.~\eqref{eq:stationary}.
Temporarily define \(a=kR_2\) and \(\zeta= r/R_2\) in the inviscid case. The characteristic equation in this case is
\(\left[\frac{\mathrm{d^{2}}}{\mathrm{d}\zeta^{2}} + \frac{1}{\zeta} \frac{\mathrm{d}}{\mathrm{d}\zeta} - \frac{1}{\zeta^{2}} - (a^2 + 8AR_{2}^{2}/\nu_o)  \right]u = 0\) with boundary conditions \(u=0\) at \(\zeta=\eta\) and \(\zeta=1\).
To have a nontrivial solution of \(u\), there must be \(a^2 + 8AR_{2}^{2}/\nu_o < 0\) and \(J_1(\beta \eta) Y_1(\beta) = J_1(\beta) Y_1(\beta\zeta)\), where we have denoted $\beta = \sqrt{|a^2 + 8AR_{2}^{2}/\nu_o|}$.
One can see the critical wave-number vanishes $a_{c}=0$.
Denote the lowest solution of \(\beta\) as \(\beta_1\) and define \(\omega_1 = \Omega_1 R_1^2/\nu\), \(\omega_2 = \Omega_2 R_2^2/\nu\) and  \(\bar{\nu}_{\text{oc}} = 1/[\beta_1^2 (1-\eta)^2 (\eta^{-2}-1)]\).
Then the stable condition is
\begin{align}
\label{eq:critical-line-inviscid}
\frac{\omega_1}{\omega_2}
\begin{cases}
< \frac{1}{1 - \bar{\nu}_o /\bar{\nu}_{\text{oc}}}, & 0<\bar{\nu}_o  < \bar{\nu}_{\text{oc}}, \\
< \frac{1}{1 - \bar{\nu}_o /\bar{\nu}_{\text{oc}}} \text{ or } > 0, & \bar{\nu}_o > \bar{\nu}_{\text{oc}}, \\
> \frac{1}{1 - \bar{\nu}_o /\bar{\nu}_{\text{oc}}} \text{ or } < 0, & \bar{\nu}_o < 0.
\end{cases}
\end{align}

Shown in Fig.~\ref{fig:phase-diagram-inviscid} are the regions of stability and instability in the \((\omega_2,\,\omega_1)\) plane. When \(0 < \bar{\nu}_{o} < \bar{\nu}_{\text{oc}}\) [Fig.~\ref{fig:phase-diagram-inviscid} (a)], the region of stability extends above the Rayleigh criterion \(\omega_1 = \omega_2\) without odd viscosity, indicating that positive \(\bar{\nu}_o\) promotes stability and suppresses instability. When \(\bar{\nu}_o\) further increases to \(\bar{\nu}_o > \bar{\nu}_{\text{oc}}\) [Fig.~\ref{fig:phase-diagram-inviscid} (b)], the critical line strides over the \(\omega_2 = 0\) axis. As long as the angular velocities of the inner and outer cylinders \(\Omega_1,\, \Omega_2\) have the same sign, the flow will always be stable no matter what their magnitudes. When the signs of \(\Omega_1\) and \(\Omega_2\) differ, the instability kicks in only when \(|\Omega_2|\) is large enough. When \(\bar{\nu}_o < 0\), the critical line falls below the Rayleigh criterion, and the region of stability and instability are swapped compared to the case \(\bar{\nu}_o \geq 0\). The flow is always stable when \(\Omega_1\) and \(\Omega_2\) have different signs, becoming unstable only when they have the same signs and the \(\Omega_2\) is large enough for fixed \(\Omega_1\). We will see in what follows that the critical lines for \(\bar{\nu}> 0\) act as {\sl limiting lines} for the viscous case. However, when \(\nu\) is turned on, the stability region for \(\bar{\nu}_o < 0\) will be reversed, as will be shown latter.

\paragraph{Narrow gap.}
\begin{figure}[t]
  \centering
  \includegraphics[width=0.99\linewidth]{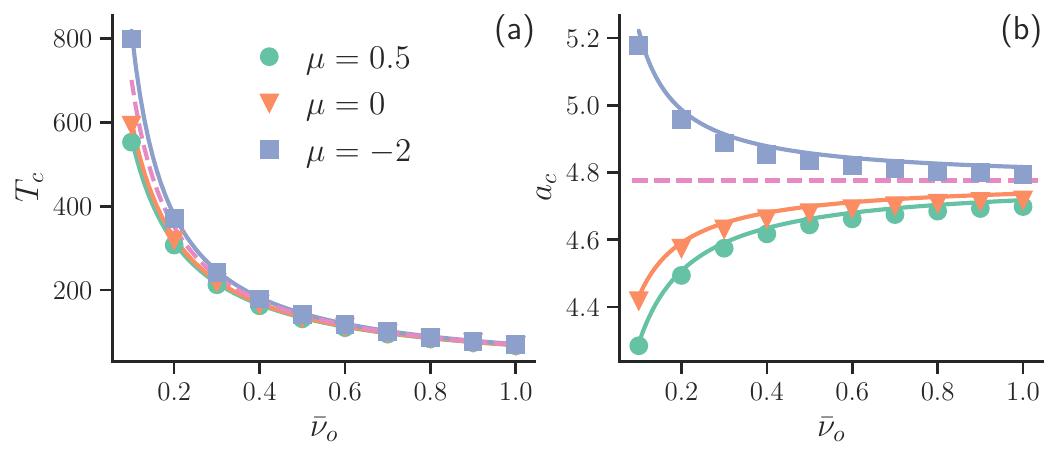}
  \caption{\label{fig:narrow-Tc-ac}Critical Taylor number $T_c$ (a) and critical wave-number $a_c$ (b) versus the reduced odd viscosity $\bar{\nu}_o$ for different $\mu=\Omega_2/\Omega_1$ in the narrow-gap case. The solid lines represent analytical results obtained by the Galerkin method, while the discrete data is obtained numerically from the superposition method. The dashed lines in (a) and (b) are, respectively, the $T_c$ and $a_c$ in the limit of large $\bar{\nu}_o$.}
\end{figure}

Let us now consider the viscous case with a narrow gap
\(d = R_2 - R_1 \ll \frac{1}{2} (R_1 + R_2)\), equivalently, \(\eta\approx 1\).
To the leading order in this limit, \(D_{*}\) becomes identical to \(D\)
and the characteristic equations reduce to
\begin{subequations}
\begin{align}
\label{eq:narrow-char1}
&(D^2 -a^2 )^2 u = \left[ 1-(1-\mu)\zeta - \bar{\nu}_o (D^2-a^2) \right]v,\\
\label{eq:narrow-char2}
& (D^2 -a^2) v = - T a^2 u.
\end{align}
\end{subequations}

\begin{figure*}[ht]
  \centering
  \includegraphics[width=0.99\linewidth]{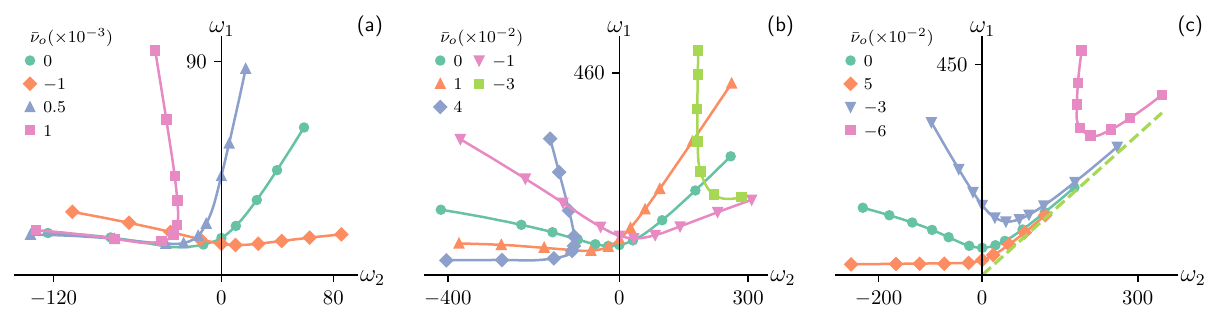}
  \caption{\label{fig:phase-diagram}Stability diagrams in the $(\omega_2,\, \omega_1)$ plane of the viscous case for different values of the reduced odd viscosity $\bar{\nu}_o$ and gap size $\eta$. The solid lines represent interpolations of the discrete data. (a) $\eta=0.1$, (b) $\eta=0.5$ and (c) narrow gap case $\eta\to 1$. The dashed line is the Rayleigh criterion.}
\end{figure*}

To solve the above characteristic equations, we first use the Galerkin method by following~\cite{chandrasekhar1961}.
Note \(v=0\) at \(\zeta = 0\) and \(\zeta=1\) can be expanded into a Fourier series
\(v = \sum_{m=1}^{\infty} C_m \sin m\pi \zeta\), where \(C_m\) are as yet undetermined coefficients. Substituting \(v\) into Eq.~\eqref{eq:narrow-char1}, solving \(u\) and then substituting both \(u\) and \(v\) in the form of the Fourier series into Eq.~\eqref{eq:narrow-char2} lead to the secular equation, from which,
to the leading order, we get the Taylor number
\begin{align}
T(a;\mu,\bar{\nu}_o) = \frac{2\, h(a)}{1+\mu + 2\bar{\nu}_o(a^2 + \pi^2)},
\end{align}
where \(h(a)= \frac{(a^2+\pi^2)^3}{a^2 \left\{ 1 - 16 a \pi^2 \cosh^2 \frac{a}{2} / \left[ (a^2+\pi^2)^2 \left( a + \sinh a \right) \right] \right\}}\).
If \(\bar{\nu}_o = 0\), the above \(T\) is reduced exactly to the case  without odd viscosity~\cite{chandrasekhar1961}.
One can see that when \(\bar{\nu}_o\) is positive and large, \(T\) can be approximated by
\(T'(a;\bar{\nu}_o) = h(a)/[\bar{\nu}_o(a^2 + \pi^2)]\).
In this case, the critical wave-number \(a_c \approx 4.776\) is independent of \(\mu\) and \(\bar{\nu}_o\).
By substitution of \(a_c\) back, the critical Taylor number can be approximated
by a universal expression \(T_c(\bar{\nu}_o) = T'(a_c;\bar{\nu}_o)\) which is independent of \(\mu\).

The Galerkin method relies on \(u\) and \(v\) being well approximated by some of the lower order basis functions.
Otherwise, higher order terms are required and the calculations quickly get very complicated.
To obtain more accurate solutions of the characteristic equations in general situations,
we use the superposition method~\cite{na1979,harrisStabilityViscousFlow1964} to transform the boundary value problem into an initial value problem and numerically solve it. Define
\(U = v\), \(V = D v\), \(W = u\), \(X= D u\), \(Y = (D^2 - a^2) u\) and \(Z = D(D^2 - a^2) u\).
The characteristic equations then reduce to a system of first order equations
\begin{align}
& DZ - a^2 Y = [1-(1-\mu)\zeta] U - \bar{\nu}_o (DV - a^2 U), \nonumber\\
& DV - a^2 U = - Ta^2 W, \quad DU = V, \\
& DW = X,\quad DX - a^2 W = Y, \quad D Y = Z.\nonumber
\end{align}
In order to satisfy the boundary conditions, the initial conditions are now stipulated 
\(U = W = X = 0 \text{ at } \zeta = 0\). For \(V\), \(Y\) and \(Z\), we impose three different initial conditions and denote the  corresponding solutions of the equations as \(U_i\), \(V_{i}\), \(W_i\), \(X_i\), \(Y_i\), \(Z_i\), \(i=1,2,3\).
The three different initial conditions for \(V\), \(Y\) and \(Z\) are \([V_i,\, Y_i,\, Z_i]_{\zeta=0} = \mathcal{I}_{3}\) with $\mathcal{I}_{3}$ being the order $3$ identity matrix.
The solution to the original boundary value problem can be expressed by
\(U = \sum_{i=1}^3 A_{i} U_{i}\), \(W = \sum_{i=1}^3 A_{i} W_{i}\) and \(X = \sum_{i=1}^3 A_{i} X_{i}\),
where \(A_i,\,i=1,2,3\) are superposition coefficients to be determined by the boundary conditions at \(\zeta=1\).
To have a nontrivial solution of \(A_i\), there must be \(\|[U_i,\, W_i,\, X_i]\|_{\zeta=1} = 0\),
from which \(A_i\) can be obtained up to an overall constant.

Shown in Fig.~\ref{fig:narrow-Tc-ac} are the results of the critical Taylor number \(T_c\) [Fig.~\ref{fig:narrow-Tc-ac} (a)] and critical wave-number \(a_c\) [Fig.~\ref{fig:narrow-Tc-ac} (b)] versus the reduced odd viscosity \(\bar{\nu}_o\) with different \(\mu\). One can see that the analytical results from the Galerkin method (solid lines) and numerical results from the superposition method (discrete data) agree well, corroborating each other.
When the reduced odd viscosity is large, the critical wave-number approaches a constant value about \(a_c\approx 4.776\) [dashed line in Fig.~\ref{fig:narrow-Tc-ac} (b)] and the critical Taylor number can be well approximated by the universal \(T_c\approx 70/\bar{\nu}_o\) [dashed line in Fig.~\ref{fig:narrow-Tc-ac} (a)].

Shown in Fig.~\ref{fig:phase-diagram} (c) are the regions of stability and instability in the \((\omega_2,\, \omega_1)\) plane for different $\bar{\nu}_{o}$ in the narrow gap case.
One can see overall positive \(\bar{\nu}_o\) lowers the critical lines, indicating it promotes the instability, while negative \(\bar{\nu}_o\) raises the critical lines and hence suppresses the instability.
These effects of the odd viscosity present themselves most obviously 
in the left half plane, where \(\Omega_2\) and \(\Omega_1\) have different signs.
In the right half plane, the critical lines with different \(\bar{\nu}_o\) all approaches the Rayleigh criterion in the large \(\omega_2\) limit.

\paragraph{Wide gap.}
Just as in the narrow gap case one can transform the boundary value problem in the wide gap case [Eqs.~(\ref{eq:wide-char1}) and (\ref{eq:wide-char2})] into an initial value problem by the method of superposition~\cite{suppl}.
The stability diagrams in the wide gap case with \(\eta = 0.1\) and \(\eta=0.5\) are shown in Figs.~\ref{fig:phase-diagram} (a) and (b), respectively.

From Fig.~\ref{fig:phase-diagram}, one can see that the reduced odd viscosity \(\bar{\nu}_o\) has a  ``lever'' effect on the critical lines.
When \(\bar{\nu}_o\) increases, the right-hand side is lifted up and the left-hand side is pushed down. When \(\bar{\nu}_o\) decreases, the opposite happens, \emph{i.e.}, the right-hand side is pushed down and the left-hand side is lifted up.
When \(\bar{\nu}_o>0\) is sufficiently large, the critical lines are completely excluded from the right half plane [Figs.~\ref{fig:phase-diagram} (a) and (b)],
indicating that the instability is completely suppressed as long as \(\Omega_1\Omega_2>0\), or \(\Omega_1\) is sufficiently large or small when \(\Omega_1\Omega_2<0\).
When \(\bar{\nu}_o < 0\) and \(|\bar{\nu}_o|\) is sufficiently large, the critical lines are completely excluded from the left half plane [Figs.~\ref{fig:phase-diagram} (b) and (c)], indicating that the instability is completely suppressed as long as \(\Omega_1\Omega_2 <0\), or \(\Omega_1\) is sufficiently large or small when \(\Omega_1\Omega_2 > 0\).
The ``lever'' effect is more prominent in both half planes for wider gap [Fig.~\ref{fig:phase-diagram} (a)],
while it is diminished in the right half plane for the narrow gap limit [Fig.~\ref{fig:phase-diagram} (c)] due to the Rayleigh criterion acting as a barrier.

It is noteworthy that the critical lines from the inviscid case [see Fig.~\ref{fig:phase-diagram-inviscid}] act as limits for the viscous case when \(\bar{\nu}_o > 0\).
However, when \(\bar{\nu}_o < 0\), not only are the stability regions of the inviscid case reversed, but the critical lines can extend across the inviscid limit. As the inviscid limit is below the Rayleigh criterion for \(\bar{\nu}_o < 0\),
the critical line goes through the Rayleigh criterion (\(\mu = \eta^2\)) and the Taylor number turns negative. In this case, the critical Taylor number \(T_c\) and wave-number \(a_c\) are obtained by maximizing \(T(a)\) instead of minimizing it.

\begin{figure}[t]
  \centering
  \includegraphics[width=0.95\linewidth]{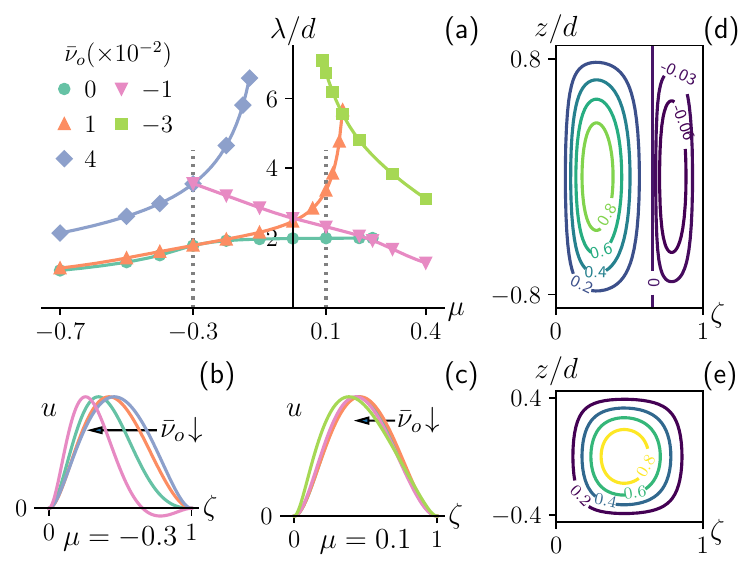}
  \caption{\label{fig:streamline}Characterizations of the Taylor vortices at the onset of instability with gap size $\eta=0.5$. (a) The wavelengths $\lambda$ as a function of $\mu$ for different reduced odd viscosities $\bar{\nu}_o$. The solid lines represent interpolations of the discrete data. (b,c) The radial velocity profiles along the gap $u(\zeta)$ for, respectively, $\mu=-0.3$ and $\mu=0.1$ and different reduced odd viscosities $\bar{\nu}_o$. (d,e) The cell patterns described by the stream functions for, respectively, $\bar{\nu}_o = -0.01$, showing a secondary vortex near the outer surface, and $\bar{\nu}_o = 0.01$ with $\mu=-0.3$ in both panels.}
\end{figure}

The Taylor vortices at the onset of the instability can be characterized by the wavelengths [Fig.~\ref{fig:streamline} (a)], radial velocity profiles [Fig.~\ref{fig:streamline} (b) and (c)] and cell patterns quantified by the stream functions [Fig.~\ref{fig:streamline} (d) and (e)].
One can observe from Fig.~\ref{fig:streamline} (a) that when the instability is suppressed, the wavelength generally grows larger irrespective of the signs of \(\mu\) and \(\bar{\nu}_o\).
When \(\bar{\nu}_o < 0\), the wavelength can fall well below the value without odd viscosity as $\mu$ increases, see line \(\bar{\nu}_o = -0.01\) in Fig.~\ref{fig:streamline} (a).
For fixed \(\mu\), the centers of the vortices indicated by the local stationary points of the radial velocity profiles \(u(\zeta)\)
are dragged inwards as the odd viscosity decreases [Fig.~\ref{fig:streamline} (b) and (c)].
In addition, for \(\mu=-0.3\) and \(\bar{\nu}_o=-0.01\), one sees the emergence of {\em secondary vortices} near the outer cylinder surface [Figs.~\ref{fig:streamline} (b) and (d)].
The stream functions describing the cell patterns in Fig.~\ref{fig:streamline} (d) and (e) are \(\sim (\eta+\zeta) u(\zeta)\cos (z/\lambda)\), being normalized so that the maximum value is \(1\).

\paragraph{Discussion and conclusions.}
Notably, the presence of odd viscosity \(\nu_o\) along \(z\)-direction does not alter the torque applied on a non-moving outer cylinder ($\Omega_2=0$) in the laminar flow,
\(\tau = 4\pi\, \rho\, \nu H \frac{R_1^2 R_2^2}{R_2^2 - R_1^2} \Omega_1\), so it can still be used in experiments to indicate the start of the instability~\cite{drazinHydrodynamicStability2004,chandrasekhar1961}.

It is of course natural to consider next the scenario where the odd viscosity vector does not align with the axisymmetric axis of the flow, which would introduce complex coefficients in the characteristic equations.
In this scenario, some of the legitimate questions to ask would be how the stationary flow, the torque on the cylinders, the stability, the principle of the  exchange of stabilities, \emph{etc}., are changed, the answers to which we leave for future endeavors.

To sum up, we have investigated how the presence of odd viscosity changes the stability of the laminar Taylor-Couette flow under axisymmetric perturbations by the linear stability analysis.
In the narrow gap case, we obtained both numerical and analytical results that perfectly match. In the more general wide gap case, we employed extensive numerical calculations and
found that the odd viscosity exerts a ``lever'' effect on the critical lines in the stability diagram. The otherwise unstable flow with high adverse distribution of angular velocity under axisymmetric disturbances can be stabilized by the odd viscosity irrespective of its sign. It remains a possible area for future research to investigate if such conclusions still hold within the nonlinear stability analysis or under non-axisymmetric disturbances.
Actually, the presence of odd viscosity probably changes a series of instabilities and transitions beyond the simple form of the Taylor vortex instability in Taylor-Couette geometry, awaiting further investigations both theoretically as well as experimentally.

\begin{acknowledgments}
G.D.~and R.P.~acknowledge the funding for the Key Project of the National Natural Science Foundation of China (NSFC) (Grant No.~12034019) and the support by the Fundamental Research Funds for the Central Universities (Grant No.~E2EG0204).
\end{acknowledgments}

%

\ifarXiv
    

\section*{Supplementary material}

# Stability of Taylor-Couette Flow with Odd Viscosity

# Supplemental Material

Guangle Du[1] and Rudolf Podgornik[1, 2, 3, 4, *]

[1]*School of Physical Sciences, University of Chinese Academy of Sciences, Beijing 100049, China*
[2]*Kavli Institute for Theoretical Sciences, University of Chinese Academy of Sciences, Beijing 100049, China*
[3]*Wenzhou Institute, University of Chinese Academy of Sciences, Wenzhou, Zhejiang 325001, China*
[4]*Beijing National Laboratory for Condensed Matter Physics and Laboratory of Soft Matter Physics,
Institute of Physics, Chinese Academy of Sciences, Beijing 100190, China*
(Dated: September 21, 2023)

The following sections are included in this Supplemental Material: derivation of the Navier-Stokes equations in cylindrical coordinates (Sec. I), derivations of perturbation equations and characteristic equations (Sec. II), solution of characteristic equations in the wide gap case by superposition method (Sec. III), numerical support of the principle of the exchange of stabilities with odd viscosity (Sec. IV), and solution of characteristic equations in the wide gap case by Galerkin method (Sec. V).

## I. DERIVATION OF THE NAVIER-STOKES EQUATIONS IN CYLINDRICAL COORDINATES

To derive the Navier-Stokes (NS) equations in cylindrical coordinates, we start from the covariant form of NS equations in general curvilinear coordinates

$$\rho(\dot{v}^i + v^j v^i_{;j}) = \sigma^{ij}_{;j}, \tag{S1}$$

where $\rho$ is the density, $v^i$ is the velocity field and semicolon represents covariant derivative. $\sigma^{ij}$ is the stress tensor

$$\sigma^{ij} = -pg^{ij} + \left(\eta_e^{ijkl} + \eta_o^{ijkl}\right) v_{k;l}, \tag{S2}$$

where $p$ is the pressure, $g^{ij}$ is the metric tensor and $\eta_e^{ijkl}$ and $\eta_o^{ijkl}$ are, respectively, the even and odd viscosity tensors that can be written explicitly as [1]

$$\eta_e^{ijkl} = \lambda g^{ij} g^{kl} + \eta(g^{ik} g^{jl} + g^{il} g^{jk}), \tag{S3a}$$

$$\eta_o^{ijkl} = -\frac{1}{4} l_n \left(\epsilon^{nik} g^{jl} + \epsilon^{njl} g^{ik} + \epsilon^{nil} g^{jk} + \epsilon^{njk} g^{il}\right). \tag{S3b}$$

$\lambda$ is the volume viscosity, $\eta$ is the even dynamical viscosity, $l_n \equiv l^m g_{mn}$ is the odd dynamical viscosity vector, and $\epsilon^{ijk}$ is alternating tensor related to Levi-Civita symbol $\varepsilon^{ijk}$ by $\epsilon^{ijk} = \varepsilon^{ijk}/\sqrt{g}$, $\epsilon_{ijk} = \sqrt{g}\,\varepsilon_{ijk}$ with $g$ being the determinant of metric tensor $g = \det(g_{ij})$. Note that in the main text, we used $\lambda$ and $\eta$ to denote, respectively, wavelength and the ratio of the two cylinder radii. We abused the notation a little here, since the chance to introduce ambiguity in this context is low. Levi-Civita connection in a Riemannian manifold is compatible with metric tensor, i.e., $g^{ij}_{;k} = 0$. Assume $l_n$ is constant with respect to covariant derivative $l_{n;j} = 0$. Then NS equations become

$$\rho(\dot{v}^i + v^j v^i_{;j}) = -g^{ij} p_{;j} + \left(\eta_e^{ijkl} + \eta_o^{ijkl}\right) v_{k;lj}. \tag{S4}$$

It can be proved that the following identity holds

$$\epsilon^{nik} g^{jl} v_{k;lj} = \epsilon^{nik} v^j_{\ ;kj} - \epsilon^{ipq} v_{q;p}{}^n + \epsilon^{npq} v_{q;p}{}^i. \tag{S5}$$

By using the above identity, the NS equations reduce to

$$\rho(\dot{v}^i + v^j v^i_{;j}) = -\tilde{p}^{;i} + \eta \left(v^{i;j}{}_j + v^{j;i}{}_j\right)$$
$$- \frac{1}{4} l_n \left[2\epsilon^{nik} v^j_{\ ;kj} - \epsilon^{ijk} v_{k;j}{}^n + \epsilon^{njk} \left(v_k{}^{;i}{}_j + v_{k;j}{}^i\right) + \epsilon^{njk} v^i_{\ ;kj}\right], \tag{S6}$$

---

* To whom correspondence should be addressed: rudolfpodgornik@ucas.ac.cn; Also affiliated with Department of Physics, Faculty of Mathematics and Physics, University of Ljubljana, 1000 Ljubljana, Slovenia.

where $\tilde{p} = p - \lambda\, v^j_{;j}$.

In cylindrical coordinates, the two indices of the second mixed covariant derivative commute, implying

$$\left(\epsilon^{nik} g^{jl} + \epsilon^{njl} g^{ik} + \epsilon^{nil} g^{jk} + \epsilon^{njk} g^{il}\right) v_{k;lj} = 2\epsilon^{nik} v^j_{;kj} - \epsilon^{ijk} v_{k;j}{}^n + 2\,\epsilon^{njk} v_{k;j}{}^i. \tag{S7}$$

The NS equations become

$$\rho(\dot{v}^i + v^j v^i_{;j}) = -\tilde{p}^{;i} + \eta\, v^{i;j}{}_j + \frac{1}{2} l^n \omega^i_{;n} - \frac{1}{2} l_n \epsilon^{nik} v^j_{;kj}, \tag{S8}$$

where $\tilde{p} = p - (\lambda + \eta)\, v^j_{;j} + l^j \omega_j$ is an effective pressure and $\omega^i = \epsilon^{ijk} v_{k;j}/2$ is the vorticity vector. By further applying the incompressibility condition $v^i_{;i} = 0$, we have

$$\rho(\dot{v}^i + v^j v^i_{;j}) = -\tilde{p}^{;i} + \eta\, v^{i;j}{}_j + \frac{1}{2} l^n \omega^i_{;n}, \tag{S9}$$

where $\tilde{p} = p + l^j \omega_j$.

To further simplify the problem, we assume that only $z$-component of the odd viscosity is nonvanishing. Denote the kinematic version of the only odd viscosity as $\nu_o = l^z/\rho$, the even kinematic viscosity as $\nu$, and the radial, azimuthal and axial components of the velocity, respectively, as $u_r$, $u_\theta$ and $u_z$. The vorticity vector components can be explicitly written as $\omega_r = \left(\partial_\theta u_z/r - \partial_z u_\theta\right)/2$, $\omega_\theta = \left(\partial_z u_r - \partial_r u_z\right)/2$ and $\omega_z = \left(\partial_r u_\theta + u_\theta/r - \partial_\theta u_r/r\right)/2$. After some calculations, we obtain the explicit NS equations

$$D_t u_r - \frac{u_\theta^2}{r} = \nu\left(\Delta^* u_r - \frac{2\partial_\theta u_\theta}{r^2}\right) + \frac{\nu_o}{2} \partial_z \omega_r - \frac{\partial_r \tilde{p}}{\rho}, \tag{S10a}$$

$$\left(D_t + \frac{u_r}{r}\right) u_\theta = \nu\left(\Delta^* u_\theta + \frac{2\partial_\theta u_r}{r^2}\right) + \frac{\nu_o}{2} \partial_z \omega_\theta - \frac{\partial_\theta \tilde{p}}{\rho\, r}, \tag{S10b}$$

$$D_t u_z = \nu \Delta u_z + \frac{\nu_o}{2} \partial_z \omega_z - \frac{\partial_z \tilde{p}}{\rho}, \tag{S10c}$$

where $\tilde{p} = p + l^z \omega_z$ and we have made the denotations $D_t = \partial_t + u_r \partial_r + u_\theta \partial_\theta/r + u_z \partial_z$, $\Delta = \partial_r^2 + \partial_r/r + \partial_\theta^2/r^2 + \partial_z^2$ and $\Delta^* = \Delta - 1/r^2$.

## II. DERIVATIONS OF PERTURBATION EQUATIONS AND CHARACTERISTIC EQUATIONS

In this section, we give detailed derivations of the perturbation equations and characteristic equations in both inviscid and viscous cases. Starting from the NS equations and the steady solution [Eqs. (1) and (2)] in the main text, we assume the disturbances are characterized by $(u_r, u_\theta, u_z) = (u'_r, V(r) + u'_\theta, u'_z)$, and $\tilde{p} = \tilde{P}(r) + \tilde{p}'$ and then get the perturbation equations

$$\left(\partial_t + \frac{V}{r}\partial_\theta\right) u'_r - \frac{2V u'_\theta}{r} = \nu\left(\Delta^* u'_r - \frac{2\partial_\theta u'_\theta}{r^2}\right) + \frac{\nu_o}{2} \partial_z \omega'_r - \frac{\partial_r \tilde{p}'}{\rho}, \tag{S11a}$$

$$\left(u'_r \partial_r + \frac{u'_r}{r}\right) V(r) + \left(\partial_t + \frac{V}{r}\partial_\theta\right) u'_\theta = \nu\left(\Delta^* u'_\theta + \frac{2\partial_\theta u'_r}{r^2}\right) + \frac{\nu_o}{2} \partial_z \omega'_\theta - \frac{\partial_\theta \tilde{p}'}{\rho\, r}, \tag{S11b}$$

$$\left(\partial_t + \frac{V}{r}\partial_\theta\right) u'_z = \nu \Delta u'_z + \frac{\nu_o}{2} \partial_z \omega'_z - \frac{\partial_z \tilde{p}'}{\rho}, \tag{S11c}$$

where $\omega'_r = \left(\partial_\theta u'_z/r - \partial_z u'_\theta\right)/2$, $\omega'_\theta = \left(\partial_z u'_r - \partial_r u'_z\right)/2$, and $\omega'_z = \left(\partial_r u'_\theta + u'_\theta/r - \partial_\theta u'_r/r\right)/2$. Under axisymmetric disturbances, the above perturbation equations reduce to

$$\partial_t u'_r - \frac{2V u'_\theta}{r} = \nu \Delta^* u'_r + \frac{\nu_o}{2} \partial_z \omega'_r - \frac{\partial_r \tilde{p}'}{\rho}, \tag{S12a}$$

$$\partial_t u'_\theta + \left(\partial_r V + \frac{V}{r}\right) u'_r = \nu \Delta^* u'_\theta + \frac{\nu_o}{2} \partial_z \omega'_\theta, \tag{S12b}$$

$$\partial_t u'_z = \nu \Delta u'_z + \frac{\nu_o}{2} \partial_z \omega'_z - \frac{\partial_z \tilde{p}'}{\rho}, \tag{S12c}$$

where $\omega'_r = -\partial_z u'_\theta/2$, $\omega'_\theta = (\partial_z u'_r - \partial_r u'_z)/2$, and $\omega'_z = (\partial_r u'_\theta + u'_\theta/r)/2$. The equation of continuity becomes $(\partial_r + 1/r) u'_r + \partial_z u'_z = 0$.

By performing the standard normal mode analysis $(u'_r, u'_\theta, u'_z) = (u(r), v(r), w(r))e^{st+ikz}$ and $\tilde{p}' = q(r)e^{st+ikz}$, we obtain the characteristic equations

$$su - 2\Omega v = \nu(DD_* - k^2)u + \frac{\nu_o}{4}k^2 v - \frac{\partial_r q}{\rho}, \tag{S13a}$$

$$sv + uD_*V = \nu(DD_* - k^2)v + \frac{\nu_o}{4}(-k^2 u - ik\partial_r w), \tag{S13b}$$

$$sw = \nu(D_*D - k^2)w + \frac{\nu_o}{4}ikD_*v - ik\frac{q}{\rho}, \tag{S13c}$$

$$0 = D_*u + ikw, \tag{S13d}$$

where we have introduced $D = d/dr$ and $D_* = d/dr + 1/r$. Elimination of $w$ and $q/\rho$ reduces the characteristic equation into

$$\frac{1}{k^2}\left[\nu(DD_* - k^2) - s\right](DD_* - k^2)u + \left[\frac{\nu_o}{4}(DD_* - k^2) - 2\Omega\right]v = 0, \tag{S14a}$$

$$\left[\frac{\nu_o}{4}(DD_* - k^2) - D_*V\right]u + \left[\nu(DD_* - k^2) - s\right]v = 0. \tag{S14b}$$

By setting $\nu = 0$ and assuming the marginal state is stationary $s = 0$, we have the characteristic equations for the inviscid case

$$\left[\frac{\nu_o}{4}(DD_* - k^2) - D_*V\right]u = 0. \tag{S15}$$

Note that since there is no boundary condition for $v$, only the equation for $u$ is left. Consider the azimuthal velocity $V = Ar + B/r$ and define $\zeta = r/R_2$ and $a = kR_2$. The characteristic equation then becomes

$$\left[\frac{d^2}{d\zeta^2} + \frac{1}{\zeta}\frac{d}{d\zeta} - \frac{1}{\zeta^2} - \left(a^2 + \frac{8AR_2^2}{\nu_o}\right)\right]u = 0 \tag{S16}$$

with the boundary conditions $u = 0$ at $\zeta = \eta$ and $\zeta = 1$.

For the viscous case, by defining $\zeta = (r - R_1)/d$, $a = kd$ and $\sigma = sd^2/\nu$, the characteristic equations Eq. (S14) can be written explicitly as

$$(DD_* - a^2 - \sigma)(DD_* - a^2)u\frac{\nu}{2\Omega_1 a^2 d^2} = \left\{\left[\frac{\mu - \eta^2}{1 - \eta^2} + \frac{1 - \mu}{1 - \eta^2}\frac{1}{\left(1 + \zeta\frac{1-\eta}{\eta}\right)^2}\right] - \frac{\nu_o}{8\Omega_1 d^2}(DD_* - a^2)\right\}v, \tag{S17a}$$

$$(DD_* - a^2 - \sigma)v = \left[1 - \frac{\nu_o}{8Ad^2}(DD_* - a^2)\right]u\frac{2Ad^2}{\nu}, \tag{S17b}$$

where now $D = d/d\zeta$ and $D_* = d/d\zeta + 1/[\zeta + \eta/(1 - \eta)]$. By substituting $u\nu/(2\Omega_1 a^2 d^2) \to u$ and defining the Taylor number and the reduced odd viscosity

$$T \equiv -\frac{4A\Omega_1 d^4}{\nu^2}, \quad \bar{\nu}_o \equiv \frac{\nu_o}{8\Omega_1 d^2}, \tag{S18}$$

we obtain the characteristic equations for the wide gap case

$$(DD_* - a^2 - \sigma)(DD_* - a^2)u = \left\{\left[\frac{\mu - \eta^2}{1 - \eta^2} + \frac{1 - \mu}{1 - \eta^2}\frac{1}{\left(1 + \zeta\frac{1-\eta}{\eta}\right)^2}\right] - \bar{\nu}_o(DD_* - a^2)\right\}v, \tag{S19a}$$

$$(DD_* - a^2 - \sigma)v = -Ta^2\left[1 - \frac{\bar{\nu}_o}{\frac{\mu - \eta^2}{1 - \eta^2}}(DD_* - a^2)\right]u. \tag{S19b}$$

The no-slip boundary conditions stipulate that $u = v = w = 0$. Due to $D_*u + ikw = 0$, we have $Du = 0$, implying that the complete boundary conditions for $u$ and $v$ are $u = v = Du = 0$ at $\zeta = 0$ and $\zeta = 1$.

To the leading order of the narrow gap approximation $1 - \eta \ll 0$, there are

$$\frac{\mu - \eta^2}{1 - \eta^2} + \frac{1 - \mu}{1 - \eta^2} \frac{1}{\left(1 + \zeta \frac{1-\eta}{\eta}\right)^2} \approx \frac{\mu - \eta^2}{1 - \eta^2} + \frac{1 - \mu}{1 - \eta^2} \left(1 - 2\zeta \frac{1-\eta}{\eta}\right) = 1 - (1 - \mu)\zeta, \tag{S20}$$

$\bar{\nu}_o(1-\eta^2)/(\mu-\eta^2) \to 0$, and $D_* \approx D$ due to that $1/[\zeta + \eta/(1-\eta)] \to 0$ in $D_*$. So the characteristic equations in the narrow gap case are

$$\left(D^2 - a^2 - \sigma\right)\left(D^2 - a^2\right) u = \left[1 - (1-\mu)\zeta - \bar{\nu}_o\left(D^2 - a^2\right)\right] v, \tag{S21a}$$

$$\left(D^2 - a^2 - \sigma\right) v = -Ta^2 u. \tag{S21b}$$

## III. SOLUTION OF CHARACTERISTIC EQUATIONS IN THE WIDE GAP CASE BY SUPERPOSITION METHOD

Similar with the narrow gap case shown in the main text, one can transform the boundary value problem in the wide gap case into an initial value problem by the method of superposition. Define

$$U = v, \quad V = D_* v, \quad W = u,$$
$$X = D_* u, \quad Y = \left(DD_* - a^2\right) u, \quad Z = D_*\left(DD_* - a^2\right) u. \tag{S22}$$

Then we have a system of first order linear equations

$$DZ - a^2 Y = \left[\frac{\mu - \eta^2}{1 - \eta^2} + \frac{1 - \mu}{1 - \eta^2}\left(1 + \frac{1-\eta}{\eta}\zeta\right)^{-2}\right] U - \bar{\nu}_o\left(DV - a^2 U\right), \tag{S23a}$$

$$DV - a^2 U = -Ta^2 W - Ta^2 \bar{\nu}_o \frac{1-\eta^2}{\eta^2 - \mu} Y, \tag{S23b}$$

$$D_* U = V, \quad D_* W = X, \quad DX - a^2 W = Y, \quad D_* Y = Z. \tag{S23c}$$

In order to satisfy the left boundary condition, there must be the following initial conditions

$$U = W = X = 0 \text{ at } \zeta = 0. \tag{S24}$$

For $V$, $Y$ and $Z$, we impose three different initial conditions and denote corresponding solutions of the equations as $U_i$, $V_i$, $W_i$, $X_i$, $Y_i$, $Z_i$, $i = 1, 2, 3$. The three different initial conditions for $V$, $Y$ and $Z$ are $[V_i, Y_i, Z_i]_{\zeta=0} = \mathcal{I}_3$ with $\mathcal{I}_3$ being the order 3 identity matrix. The solution to the original boundary value problem can be expressed by

$$U = \sum_{i=1}^{3} A_i U_i, \, W = \sum_{i=1}^{3} A_i W_i, \text{ and } X = \sum_{i=1}^{3} A_i X_i, \tag{S25}$$

where $A_i$, $i = 1, 2, 3$ are superposition coefficients to be determined by the boundary conditions at $\zeta = 1$. To have a nontrivial solution of $A_i$, there must be

$$\begin{vmatrix} U_1 & W_1 & X_1 \\ U_2 & W_2 & X_2 \\ U_3 & W_3 & X_3 \end{vmatrix}_{\zeta=1} = 0. \tag{S26}$$

The superposition coefficients $A_i$ can be obtained from the above singular matrix up to an overall constant.

## IV. NUMERICAL SUPPORT OF THE PRINCIPLE OF THE EXCHANGE OF STABILITIES WITH ODD VISCOSITY

In the main text, we assumed that the principle of the exchange of stabilities holds, *i.e.*, that the marginal state is stationary. Here we use numerical calculations to support this assertion. If the marginal state is oscillating, the

characteristic equations become

$$\left(DD_* - a^2 - i\sigma\right)\left(DD_* - a^2\right)u = \left[\frac{\mu - \eta^2}{1 - \eta^2} + \frac{1-\mu}{1-\eta^2}\left(1 + \frac{1-\eta}{\eta}\zeta\right)^{-2} - \bar{\nu}_o\left(DD_* - a^2\right)\right]v, \tag{S27a}$$

$$\left(DD_* - a^2 - i\sigma\right)v = -Ta^2\left[1 + \bar{\nu}_o\frac{1-\eta^2}{\eta^2 - \mu}\left(DD_* - a^2\right)\right]u, \tag{S27b}$$

where $i\sigma = sd^2/\nu$ is purely imaginary. The boundary conditions are still

$$u = v = Du = 0 \text{ at } \zeta = 0 \text{ and } \zeta = 1. \tag{S28}$$

We write explicitly the real and imaginary parts of $u$ and $v$

$$u = \Re u + i\Im u, \quad v = \Re v + i\Im v. \tag{S29}$$

and extract the real and imaginary parts of the characteristic equations

$$\left(DD_* - a^2\right)^2 \Re u + \sigma\left(DD_* - a^2\right)\Im u = \left[\frac{\mu - \eta^2}{1 - \eta^2} + \frac{1-\mu}{1-\eta^2}\left(1 + \frac{1-\eta}{\eta}\zeta\right)^{-2} - \bar{\nu}_o\left(DD_* - a^2\right)\right]\Re v, \tag{S30a}$$

$$\left(DD_* - a^2\right)\Re v + \sigma\Im v = -Ta^2\left[1 + \bar{\nu}_o\frac{1-\eta^2}{\eta^2-\mu}\left(DD_* - a^2\right)\right]\Re u, \tag{S30b}$$

$$\left(DD_* - a^2\right)^2 \Im u - \sigma\left(DD_* - a^2\right)\Re u = \left[\frac{\mu - \eta^2}{1 - \eta^2} + \frac{1-\mu}{1-\eta^2}\left(1 + \frac{1-\eta}{\eta}\zeta\right)^{-2} - \bar{\nu}_o\left(DD_* - a^2\right)\right]\Im v, \tag{S30c}$$

$$\left(DD_* - a^2\right)\Im v - \sigma\Re v = -Ta^2\left[1 + \bar{\nu}_o\frac{1-\eta^2}{\eta^2-\mu}\left(DD_* - a^2\right)\right]\Im u. \tag{S30d}$$

Define

$$\begin{aligned} U &= \Re v, \quad V = D_*\Re v, \quad W = \Re u, \\ X &= D_*\Re u, \quad Y = (DD_* - a^2)\Re u, \quad Z = D_*(DD_* - a^2)\Re u, \\ L &= \Im v, \quad M = D_*\Im v, \quad N = \Im u, \\ O &= D_*\Im u, \quad P = (DD_* - a^2)\Im u, \quad Q = D_*(DD_* - a^2)\Im u. \end{aligned} \tag{S31}$$

Then we have the following first order linear equations

$$DZ - a^2 Y + \sigma P = \left[\frac{\mu - \eta^2}{1 - \eta^2} + \frac{1-\mu}{1-\eta^2}\left(1 + \frac{1-\eta}{\eta}\zeta\right)^{-2}\right]U - \bar{\nu}_o(DV - a^2 U),$$

$$DV - a^2 U + \sigma L = -Ta^2 W - Ta^2\bar{\nu}_o\frac{1-\eta^2}{\eta^2-\mu}Y,$$

$$DQ - a^2 P - \sigma Y = \left[\frac{\mu - \eta^2}{1 - \eta^2} + \frac{1-\mu}{1-\eta^2}\left(1 + \frac{1-\eta}{\eta}\zeta\right)^{-2}\right]L - \bar{\nu}_o(DM - a^2 L),$$

$$DM - a^2 L - \sigma U = -Ta^2 N - Ta^2\bar{\nu}_o\frac{1-\eta^2}{\eta^2-\mu}P,$$

$$D_*U = V, \quad D_*W = X, \quad DX - a^2 W = Y, \quad D_*Y = Z,$$

$$D_*L = M, \quad D_*N = O, \quad DO - a^2 N = P, \quad D_*P = Q. \tag{S32}$$

When $\sigma = 0$, the above equations decouple into two independent and equivalent sets of equations.

In order to satisfy the boundary conditions, there must be the following initial conditions

$$U = W = X = L = N = O = 0 \text{ at } \zeta = 0. \tag{S33}$$

For $V$, $Y$, $Z$, $M$, $P$ and $Q$, we impose six different initial conditions and denote corresponding solutions of the equations as $U_i$, $V_i$, $W_i$, $X_i$, $Y_i$, $Z_i$, $L_i$, $M_i$, $N_i$, $O_i$, $P_i$ and $Q_i$, $i = 1, 2, \cdots, 6$. The six different initial conditions for $V$, $Y$, $Z$, $M$, $P$ and $Q$ are

$$[V_i, Y_i, Z_i, M_i, P_i, Q_i]_{\zeta=0} = \mathcal{I}_6, \tag{S34}$$

where $\mathcal{I}_6$ is the order 6 identity matrix. The solution of the original boundary value problem can be expressed by

$$U = \sum_{i=1}^{6} A_i U_i, \quad W = \sum_{i=1}^{6} A_i W_i, \quad X = \sum_{i=1}^{6} A_i X_i,$$

$$L = \sum_{i=1}^{6} A_i L_i, \quad N = \sum_{i=1}^{6} A_i N_i, \text{ and } O = \sum_{i=1}^{6} A_i O_i, \tag{S35}$$

where $A_i$, $i = 1, \ldots, 6$ are superposition coefficients to be determined by the boundary conditions at $\zeta = 1$. To have a nontrivial solution of $A_i$, there must be

$$\begin{vmatrix} U_1 & W_1 & X_1 & L_1 & N_1 & O_1 \\ U_2 & W_2 & X_2 & L_2 & N_2 & O_2 \\ U_3 & W_3 & X_3 & L_3 & N_3 & O_3 \\ U_4 & W_4 & X_4 & L_4 & N_4 & O_4 \\ U_5 & W_5 & X_5 & L_5 & N_5 & O_5 \\ U_6 & W_6 & X_6 & L_6 & N_6 & O_6 \end{vmatrix}_{\zeta=1} = 0. \tag{S36}$$

We scanned the parameter space (with the third value in square brackets being step) $\sigma \in [2, 10, 2]$, $\bar{\nu}_o \in [-0.01, 0.01, 0.002]$, $a \in [1, 5, 1]$, $\mu \in (-1, 0.2, 0.2)$, $\eta = 0.5$ by continuously varying $T \in [0, 10^5]$. We found *no parameter combination* satisfying the secular equation Eq. (S36), implying that the principle of the exchange of stabilities holds in the presence of odd viscosity in the domain of parameters investigated. Numerical points in the higher dimensional parameter space are, however, sparse and a full proof of the principle would ideally be an analytical one or based on experiments and can be most appropriately investigated in the future.

## V. SOLUTION OF CHARACTERISTIC EQUATIONS IN THE WIDE GAP CASE BY GALERKIN METHOD

In this section, we adopt a different definition of Taylor number and reduced odd viscosity from the main text and solve the characteristic equations in the wide gap case by Galerkin method [2]. We start from the characteristic equations in the wide gap case.

$$\left(DD_* - k^2 - \frac{s}{\nu}\right)\left(DD_* - k^2\right)u = \left[2\left(A + \frac{B}{r^2}\right)\frac{k^2}{\nu} - \frac{\nu_o}{4}\frac{k^2}{\nu}\left(DD_* - k^2\right)\right]v, \tag{S37a}$$

$$\left(DD_* - k^2 - \frac{s}{\nu}\right)v = \left[\frac{2A}{\nu} - \frac{\nu_o}{4\nu}\left(DD_* - k^2\right)\right]u, \tag{S37b}$$

with the boundary conditions

$$u = v = Du = 0, \quad r = R_1 \text{ and } r = R_2. \tag{S38}$$

Different from the main text, where $r$ is measured by the gap width $d$, we here measure $r$ in the unit of $R_2$ and define $r' = r/R_2$, $a = kR_2$ and $\sigma = sR_2^2/\nu$. Multiplying $R_2^4$ on Eq. (S37a) and $R_2^2$ on Eq. (S37b), substituting $2AR_2^2u/\nu \to u$ and omitting prime in $r'$, we obtain

$$\left(DD_* - a^2 - \sigma\right)\left(DD_* - a^2\right)u = \left[\frac{4AR_2^4 a^2}{\nu^2}\left(A + \frac{B}{R_2^2 r^2}\right) - \frac{\nu_o}{2}\frac{AR_2^2 a^2}{\nu^2}\left(DD_* - a^2\right)\right]v, \tag{S39a}$$

$$\left(DD_* - a^2 - \sigma\right)v = \left[1 - \frac{\nu_o}{8AR_2^2}\left(DD_* - a^2\right)\right]u. \tag{S39b}$$

We now define the Taylor number, reduced odd viscosity and an additional parameter as

$$T \equiv -\frac{4ABR_2^2}{\nu^2}, \bar{\nu}_o \equiv -\frac{\nu_o}{8AR_2^2}, \text{ and } \kappa \equiv -\frac{AR_2^2}{B} = \frac{1 - \mu/\eta^2}{1 - \mu}. \tag{S40}$$

The characteristic equations now become

$$\left(DD_* - a^2 - \sigma\right)\left(DD_* - a^2\right)u = -Ta^2 \left[\frac{1}{r^2} - \kappa - \bar{\nu}_o \kappa \left(DD_* - a^2\right)\right]v, \tag{S41a}$$

$$\left(DD_* - a^2 - \sigma\right)v = \left[1 + \bar{\nu}_o \left(DD_* - a^2\right)\right]u \tag{S41b}$$

with the boundary conditions

$$u = v = Du = 0, \text{ at } r = \eta \text{ and } r = 1. \tag{S42}$$

We assume the principle of the exchange of stabilities holds, i.e., $\sigma = 0$. The characteristic equations become

$$(DD_* - a^2)^2 u = -Ta^2 \left[ \frac{1}{r^2} - \kappa - \bar{\nu}_o \kappa \left( DD_* - a^2 \right) \right] v, \tag{S43a}$$

$$(DD_* - a^2) v = \left[ 1 + \bar{\nu}_o \left( DD_* - a^2 \right) \right] u. \tag{S43b}$$

Note that the boundary conditions for $u$ are $u = Du = 0$ at $r = \eta$ and $r = 1$. To expand $u$, we employ the orthonormal and complete functions fulfilling the following characteristic value problem

$$\left( \frac{d^2}{dr^2} + \frac{1}{r}\frac{d}{dr} - \frac{\nu^2}{r^2} \right)^2 y = \alpha^4 y \tag{S44}$$

with the boundary conditions $y = 0$ and $\frac{dy}{dr} = 0$ at $r = \eta$ and $r = 1$. The general solution can be written as

$$y = AJ_\nu(\alpha r) + BY_\nu(\alpha r) + CI_\nu(\alpha r) + DK_\nu(\alpha r), \tag{S45}$$

where $\alpha$ and the coefficients $A, B, C, D$ are to be determined by the boundary conditions. Recall the recurrence relations of Bessel functions [3]

$$\frac{1}{x^\nu} \frac{d}{dx} [x^\nu Z_\nu(x)] = Z_{\nu-1}(x), \tag{S46}$$

where $Z_\nu$ can denote $J_\nu, Y_\nu, I_\nu$, and $e^{i\nu\pi} K_\nu$. Note the extra factor $e^{i\nu\pi}$ for $K_\nu$. Therefore,

$$\frac{1}{\alpha r^\nu} \frac{d}{dr}(r^\nu y) = AJ_{\nu-1}(\alpha r) + BY_{\nu-1}(\alpha r) + CI_{\nu-1}(\alpha r) - DK_{\nu-1}(\alpha r). \tag{S47}$$

The extra minus sign is from the factor $e^{i\nu\pi}$ for $K_\nu$. So the application of the boundary conditions lead to the characteristic equation

$$\begin{vmatrix} J_\nu(\alpha) & Y_\nu(\alpha) & I_\nu(\alpha) & K_\nu(\alpha) \\ J_\nu(\alpha\eta) & Y_\nu(\alpha\eta) & I_\nu(\alpha\eta) & K_\nu(\alpha\eta) \\ J_{\nu-1}(\alpha) & Y_{\nu-1}(\alpha) & I_{\nu-1}(\alpha) & -K_{\nu-1}(\alpha) \\ J_{\nu-1}(\alpha\eta) & Y_{\nu-1}(\alpha\eta) & I_{\nu-1}(\alpha\eta) & -K_{\nu-1}(\alpha\eta) \end{vmatrix} = 0. \tag{S48}$$

If $\alpha_m$ is a characteristic root and $(1, B_m, C_m, D_m)$ is the vector annihilated by the matrix, then

$$\mathscr{C}_{\nu;m}(r) = J_\nu(\alpha_m r) + B_m Y_\nu(\alpha_m r) + C_m I_\nu(\alpha_m r) + D_m K_\nu(\alpha_m r) \tag{S49}$$

is a proper solution belonging to $\alpha_m$.

We expand $u$ as

$$u(r) = \sum_{j=1}^{\infty} P_j \mathscr{C}_1(\alpha_j r) = \sum_{j=1}^{\infty} P_j \left[ u_j(r) + v_j(r) \right], \tag{S50}$$

where

$$u_j(r) = A_j J_1(\alpha_j r) + B_j Y_1(\alpha_j r), \quad v_j(r) = C_j I_1(\alpha_j r) + D_j K_1(\alpha_j r). \tag{S51}$$

Obviously, the two identities

$$(DD_* - a^2) u_j = -(\alpha_j^2 + a^2) u_j, \quad (DD_* - a^2) v_j = (\alpha_j^2 - a^2) v_j, \tag{S52}$$

imply

$$(DD_* - a^2) u = \sum_{j=1}^{\infty} P_j \left[ -(\alpha_j^2 + a^2) u_j + (\alpha_j^2 - a^2) v_j \right]. \tag{S53}$$

Eq. (S43b) becomes

$$\left(DD_* - a^2\right) v = \sum_{j=1}^{\infty} P_j \left[\left(1 - \bar{\nu}_o a^2\right)(u_j + v_j) - \bar{\nu}_o \alpha_j^2 (u_j - v_j)\right]. \tag{S54}$$

The fundamental solutions of the homogeneous equation are $I_1(ar)$ and $K_1(ar)$. A particular solution is

$$\sum_{j=1}^{\infty} P_j \left\{ -\frac{1 - \bar{\nu}_o(\alpha_j^2 + a^2)}{\alpha_j^2 + a^2} u_j + \frac{1 + \bar{\nu}_o(\alpha_j^2 - a^2)}{\alpha_j^2 - a^2} v_j \right\}. \tag{S55}$$

The general solution can then be written as

$$v = \sum_{j=1}^{\infty} P_j \left[ p_j I_1(ar) + q_j K_1(ar) + \frac{a^2(u_j + v_j) - \alpha_j^2(u_j - v_j)}{\alpha_j^4 - a^4} + \bar{\nu}_o(u_j + v_j) \right]. \tag{S56}$$

The constants of integration $p_j$ and $q_j$ are determined by the boundary conditions $v = 0$ at $r = 1$ and $r = \eta$, which leads to

$$p_j I_1(a) + q_j K_1(a) = \frac{2\alpha_j^2}{\alpha_j^4 - a^4} u_j(1),$$

$$p_j I_1(a\eta) + q_j K_1(a\eta) = \frac{2\alpha_j^2}{\alpha_j^4 - a^4} u_j(\eta). \tag{S57}$$

We have used the fact $u_j(1) = -v_j(1)$ and $u_j(\eta) = -v_j(\eta)$. Therefore,

$$p_j = \frac{2\alpha_j^2}{\Delta(\alpha_j^4 - a^4)} \left[K_1(a\eta) u_j(1) - K_1(a) u_j(\eta)\right],$$

$$q_j = \frac{2\alpha_j^2}{\Delta(\alpha_j^4 - a^4)} \left[-I_1(a\eta) u_j(1) + I_1(a) u_j(\eta)\right], \tag{S58}$$

where $\Delta = I_1(a) K_1(a\eta) - K_1(a) I_1(a\eta)$, and the right-hand side of Eq. (S43a) can be written as

$$\left(DD_* - a^2\right)^2 u = \sum_{j=1}^{\infty} P_j \left[\left(\alpha_j^2 + a^2\right)^2 u_j + \left(\alpha_j^2 - a^2\right)^2 v_j\right]$$

$$= \sum_{j=1}^{\infty} P_j \left[\left(\alpha_j^4 + a^4\right)(u_j + v_j) + 2\alpha_j^2 a^2 (u_j - v_j)\right]. \tag{S59}$$

Then Eq. (S43a) becomes

$$\sum_{j=1}^{\infty} P_j \left[\left(\alpha_j^4 + a^4\right)(u_j + v_j) + 2\alpha_j^2 a^2(u_j - v_j)\right]$$

$$= Ta^2 \left(\kappa - \frac{1}{r^2}\right) \sum_{j=1}^{\infty} P_j \left[p_j I_1(ar) + q_j K_1(ar) + \frac{a^2(u_j + v_j) - \alpha_j^2(u_j - v_j)}{\alpha_j^4 - a^4} + \bar{\nu}_o(u_j + v_j)\right]$$

$$+ \bar{\nu}_o \kappa T a^2 \sum_{j=1}^{\infty} P_j \left[\left(1 - \bar{\nu}_o a^2\right)(u_j + v_j) - \bar{\nu}_o \alpha_j^2 (u_j - v_j)\right]. \tag{S60}$$

Multiply the above equation with $r(u_k + v_k)$ and integrate over the interval $[\eta, 1]$. We first introduce some convenient

notations

$$\int_\eta^1 (u_k + v_k)(u_j + v_j) r \mathrm{d}r = N_k \delta_{kj},$$

$$M_{kj} = \int_\eta^1 (u_k + v_k)(u_j + v_j) \frac{\mathrm{d}r}{r},$$

$$\Delta_{kj}^{(\pm 1)} = \int_\eta^1 (u_k + v_k)(u_j - v_j) r^{\pm 1} \mathrm{d}r,$$

$$I_k^{(\pm 1)}(a) = \int_\eta^1 (u_k + v_k) I_1(ar) r^{\pm 1} \mathrm{d}r,$$

$$K_k^{(\pm 1)}(a) = \int_\eta^1 (u_k + v_k) K_1(ar) r^{\pm 1} \mathrm{d}r.$$

We then have

$$\sum_{j=1}^\infty P_j \left[ (\alpha_j^2 + a^2) N_k \delta_{kj} + 2\alpha_j^2 a^2 \Delta_{kj}^{(1)} \right]$$
$$= Ta^2 \sum_{j=1}^\infty P_j \Bigg\{ p_j \left[ \kappa I_k^{(1)}(a) - I_k^{(-1)}(a) \right] + q_j \left[ \kappa K_k^{(1)}(a) - K_k^{(-1)}(a) \right]$$
$$+ \left( \frac{a^2}{\alpha_j^4 - a^4} + \bar{\nu}_o \right) [\kappa N_k \delta_{kj} - M_{kj}] - \frac{\alpha_j^2}{\alpha_j^4 - a^4} \left[ \kappa \Delta_{kj}^{(1)} - \Delta_{kj}^{(-1)} \right] \Bigg\}$$
$$+ \bar{\nu}_o \kappa Ta^2 \sum_{j=1}^\infty P_j \left[ (1 - \bar{\nu}_o a^2) N_k \delta_{kj} - \bar{\nu}_o \alpha_j^2 \Delta_{jk}^{(1)} \right]. \tag{S61}$$

So the secular matrix is

$$\left[ \kappa Ta^2 \left( \frac{a^2}{\alpha_j^4 - a^4} + \bar{\nu}_o \right) + \kappa Ta^2 \bar{\nu}_o (1 - \bar{\nu}_o a^2) - (\alpha_j^2 + a^2) \right] N_k \delta_{kj} - \left[ \kappa Ta^2 \bar{\nu}_o^2 \alpha_j^2 + 2\alpha_j^2 a^2 \right] \Delta_{kj}^{(1)}$$
$$+ Ta^2 \Bigg\{ p_j \left[ \kappa I_k^{(1)}(a) - I_k^{(-1)}(a) \right] + q_j \left[ \kappa K_k^{(1)}(a) - K_k^{(-1)}(a) \right]$$
$$- \frac{\alpha_j^2}{\alpha_j^4 - a^4} \left[ \kappa \Delta_{kj}^{(1)} - \Delta_{kj}^{(-1)} \right] - \left( \frac{a^2}{\alpha_j^4 - a^4} + \bar{\nu}_o \right) M_{kj} \Bigg\}. \tag{S62}$$

The vanishing determinant of the above secular matrix leads to the neutral curve and critical Taylor number, which can only be solved numerically.

---


\fi

\end{document}